# Automatic Speech Recognition Using Template Model for Man-Machine Interface

Neema Mishra
M.Tech. (CSE) Project Student
G H Raisoni College of Engg.
Nagpur University, Nagpur,
INDIA
neema.mishra@gmail.com

Urmila Shrawankar
CSE Dept.
G H Raisoni College of Engg.
Nagpur University, Nagpur,
INDIA
urmila@ieee.org

Dr. V. M Thakare
Professor & Head
PG Dept. of Computer Science
SGB Amravati University, Amravati,
INDIA

*Abstract*
*Speech is a natural form of communication for human beings, and computers with the ability to understand speech and speak with a human voice are expected to contribute to the development of more natural man-machine interfaces. Computers with this kind of ability are gradually becoming a reality, through the evolution of speech recognition technologies. Speech being an important mode of interaction with computers. In this paper Feature extraction is implemented using well-known Mel-Frequency Cepstral Coefficients (MFCC).Pattern matching is done using Dynamic time warping (DTW) algorithm.*

## 1. Introduction

We always needed to minimize human effort to get things done. A convenient and user-friendly interface for Human Computer Interaction is an important technology issue. The prevalent computer interface is via keyboard or a pointing device for input and a visual display unit or a printer for output. Spoken languages dominate communication among human being and hence people expect speech interface with computers [1]. Automatic Speech Recognition (ASR) methods can be categories into two types; a) Text-independent (TI) and b) Text dependant (TD).Text-independent methods [2-6] assume that passwords users are uttering can be anything.TI methods not pay much attention to features dynamics and treat the sequence of extracted features from the speech utterance not as a sequence of symbols. Text-dependant speaker recognition methods [7-9] exploits the feature dynamic to capture the identity of the speaker's methods compare the features vectors sequence of the test utterance with the "feature-dynamics-model" of all the speakers.

Machine recognition of speech involves generating a sequence of words which best matches the given speech signal. In the speaker independent mode of speech recognition, the computer should ignore the specific characteristic of the speech signal and extract the intended message. To make an Automatic Speech Recognition (ASR) system, we used speech recognition by pattern matching of whole words. In this method if the same person repeats the same isolated word on separate occasion, the pattern is likely to be generally similar because the same phonetic relationships will apply. A well-established approach to ASR is to store in the machine example acoustic patterns (called template) for the words to be recognized, usually spoken by the person who will subsequently used the machine. Any incoming word is compared in turn with all words in the store and the one, which is most similar, is assumed the correct one. Exploiting this similarity is, however, critically depend on how the word patterns are compared i.e how the distance between words is calculated.

The most commonly used feature extraction methods are Linear Predictive Cepstral Coefficients (LPCC), Perceptual Linear Prediction (PLP) Cepstra, Mel-Frequency Cepstral Coefficients (MFCC).Compared to LPCC, speech recognition based on MFCC can produce higher recognition rate and suppressed noise effectively.So,MFCC featured is utilized more effectively than LPCC.Standard MFCC is very sensitive to noise for the feature of speech is static.So,here we used MFCC extraction method for better result. In this paper for pattern matching Dynamic time warping algorithm is used.

The paper organized as follows: 2.discuss about Feature Extraction 3.Pattern Recognition 4.Discuss about data collection and experimental result .5.Conclusion and future work.

## 2. Feature Extraction

The common step in feature extraction is frequency or spectral analysis. The signal processing techniques aim to extract features that are related to identify the characteristics [10].The goal of feature extraction is to find a set of properties of an utterance that have acoustic correlations in the speech signal i.e. parameters that can some how estimated through processing of signal waveform. Such parameters are termed as features [11].

Several different feature extraction algorithm exist, namely
- Linear Predictive Cepstral Coefficients(LPCC)
- Perceptual Linear Prediction(PLP) Cepstra
- Mel-Frequency Cepstral Coefficients (MFCC)

LPCC computes Spectral envelop before converting it into Cepstral coefficient. The LPCC are LP-derived cepstral coefficient. The PLP integrates critical bands, equal loudness pre emphasis and intensity to compressed loudness. The PLP is based on the Nonlinear Bark scale. The PLP is designed to speech recognition with removing of speaker dependant characteristics. MFCC are extensively in ASR.MFCC is based on signal decomposition with the help of a filter bank, which uses the Mel scale. The MFCC results on Discrete Cosine Transform (DCT) of a real logarithm of a short-time energy expressed on the Mel frequency sacle.This paper work considered 12MFCC.The Cepstral coefficients are set of feature reported to be robust in some different pattern recognition tasks concerning with human voice. Human voice is very well adapted to the ear sensitivity. In speech recognition, tasks 12 coefficients are retained which represent the slow variation of the spectrum of the signal, which characterizes the vocal tract shape of the uttered words [12].

The Mel-frequency cepstrum coefficient technique is often used to create the fingerprint of the sound files. The mfcc are based on the known variation of the human ear's critical bandwidth frequencies with filter spaced linearly at low frequencies and logarithmically at high frequencies used to capture important characteristics of speech. Studies have shown that human perception of the frequencies contents of sound for speech signal does not follow a linear sacle.Thus, for each store with an actual frequency, measures in Hz; a subjective pitch is measured on a scale called the Mel-scale. The Mel frequency scale is below 1000 Hz and a logarithmic spacing above 100Hz.As reference point the pitch of 1 KHz, tone, 40db above the perceptual hearing threshold is defined as 1000 Mels.The following formula is used to compute the Mels for particular frequency:

$$\text{Mel (f)} = 2595 * \log_{10}(1 + f/700)$$

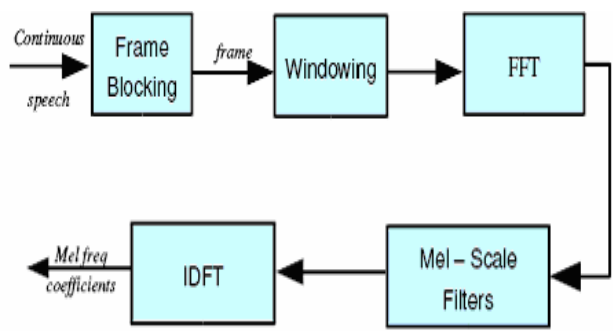

**Fig 1: Block Diagram of the MFCC Processes**

In frame, blocking the speech waveform is cropped to remove silence or acoustical interference that may be present in the beginning or end of the sound file. The windowing block minimizes the discontinuities of the signal by tapering the beginning and end of each frame to zero. The Fast Fourier Transform (FFT) block converts each frame from the time domain to the frequency domain. In the Mel Scale Filter block, the signal is filtered using band-pass filter whose bandwidths and spacing are roughly equal to critical bands and whose range of center frequencies covers frequencies most important for speech perception (300-5000Hz).In Inverse Discrete Fourier Transform (IDFT) block, Cepstrum is generated. Where Cepstrum is the spectrum of the log of the spectrum. MFCC feature is considered for speaker –independent speech recognition and for the speaker recognition task as well [13].The given table describes the comparison of LPCC and MFCC.

| Categorization | MFCC | LPCC |
|---|---|---|
| Low Bandwidth | Higher result | Less Effective |
| Noisy | Effective | Less Effective |
| Vocal Tract | Yes | No |
| Human Ear | Good | Bad |
| | | |

Table 1: Comparison between MFCC and LPCC

## 3. Pattern Recognition

The dynamic time warping is a technique of finding optimal non-linear warping of test patterns so as to obtained good match with a reference pattern(model) with a reasonable computational load [14].

In this paper, DTW based on pattern matching is done. DTW is an algorithm for measuring similarity between two sequences which may very in time or speed. DTW is a method that allow a computer to find an optimal match between two given sequences with certain restriction. In speech recognition technique the test data is converted to templates. The recognition process then consists of matching the incoming speech pattern with stored templates pattern. The template with the lowest distance measure from the input pattern is the recognized word. The optimal match (lowest distance measure) is based upon dynamic programming. This is called a Dynamic Time Warping (DTW) [1].

Let $D(i,j)$ be the cumulative distance along the optimum path from the beginning of the word to point i,j.

$$D(i, j) = \sum_{x,y=1,1}^{ij} d(x, y)$$

Along the best path

There are only three possibilities for the point before i,j it follows that
If $D(i, j)$ is the global distance up to $(i, j)$ and the local distance at $(i, j)$ is given by $d(i, j)$

**D(i,j)= min (i-1,j),D(i-1,j-1), D(i,j-1)) + d(i,j)   ……. (1)**

The value (1, 1) must be equal to d (1, 1) as this point is the beginning of all possible paths. The final global distance D (i,j) gives us the overall matching score of the template with the input. The input word is then recognized as the word corresponding to the template with the lowest matching score.

## 4. Experimental Setup

### 4.1 Data Collection

The database consists of four repetitions of ten words by 4 person (3 male and 1 female). The set of names being same for all speakers. The data colleted from the microphone contains the effect of room acoustics and the background noise in the computer room. The speech data was sampled at 16 kHz, single channel (mono) in MS wav format. The close talking microphone was used in the experiment for data collection.

### 4.2 Experiment Details

We conducted three different experiments dealing with different aspects of speech recognition system for better accuracy and practical implementation. In order to match the Test template to the reference template of the same speaker, we have record voices of speakers, converted them to sequence vectors of Mel cepstral coefficient and stored them into binary file as reference template. Now, for matching the stored reference template with the test template, we use DTW algorithm.

**Experiment 1:** Matching test and reference template of same speaker.

In this experiment first repetition of 10 words are used as the reference and next 3 repetitions of the words are used as test template. The 30(3*10) template were matched using DTW algorithm implemented in mat lab.
The accuracy for other speakers in the similar manner was:

| SPEAKER | ACCURACY |
|---|---|
| Speaker 1 | 100% |
| Speaker 2 | 96.6% |
| Speaker 3 | 96.7% |
| Speaker 4 | 96.67% |

Table 2: Result of accuracy for matching test and reference template

The average accuracy when test and reverence template of same speaker were matched is 97.5%.

**Experiment 2:** Matching reference template of one speaker with test template of another speaker.

In this experiment, first repetition of 10 names by one speaker was taken as a reference template and all templates of all others speakers was taken as test template matched against them.

For example, taking 3rd speaker as reference speaker and $2^{nd}$ speaker as test 35 out of 40 templates got matched. Therefore the accuracy is calculated as 35/40*100 =87.5%.

Here the voices being matched are different speakers, the accuracy decreased as expected. In case of mismatch of gender the accuracy decreased further (speaker 1 is female while rest are male.) All possible combination was tested and the few results are tabulated as follows:

| REFERNCE | TEST | ACCURACY |
|---|---|---|
| Speaker1 | Speaker2 | 45% |
| Speaker1 | Speaker4 | 50% |
| Speaker2 | Speaker1 | 67.5% |
| Speaker2 | Speaker3 | 80% |
| Speaker3 | Speaker4 | 80% |
| Speaker3 | Speaker1 | 85.5% |
| Speaker4 | Speaker1 | 57.5% |
| Speaker4 | Speaker3 | 87.5% |
|  |  |  |

Table 3: Result for accuracy of matching reference template of one speaker with test template of another speaker.

The average accuracy when test and reference templates of different speakers were matched was 70.6%.

## 5. CONCLUSION AND FUTURE DIRECTION

Automatic recognition of speech by machine has been a goal of research for more than five decades [15].This work proposed an approach to implement an ASR with MFCC feature extraction and for pattern matching DTW is used which match test and reference template for same speaker with accuracy 97.5%.In the other experiment for keeping matching reference template of one speaker with test template of another speaker the accuracy is 70.6%.Till now we use rectangular windowing because it is simple to understand. One could use hamming windowing to increase the accuracy rate.

Scope for future work would concentrate on incorporating the features as from limited database size to large, from speaker dependant to speaker independent. In addition, we can apply (HIDDEN MARKOV MODEL) HMM to find out the most probable word from the database, if no corresponding template is stored in the database.

## 6. REFERNCES